**Value-of-Information Analysis for External Validation of Risk Prediction Models in Multicenter Studies and Systematic Reviews**

Laure Wynants[1,2], Kim Zhipei Wang[1], Sabine Grimm[3,4], Andrea Gabrio[5], Andrew Vickers[6], Ewout Steyerberg[7], Ben Van Calster[2,7], Mohsen Sadatsafavi[8]

Affiliations:

1. Department of Epidemiology, Care and Public Health Research Institute, Maastricht University, Maastricht, The Netherlands

2. Department of Development and Regeneration, KU Leuven, Leuven, Belgium

3. Department of Clinical Epidemiology and Medical Technology Assessment, Maastricht University Medical Centre+ (MUMC+), Maastricht, The Netherlands

4. Department of Health Services Research, Care and Public Health Research Institute, Maastricht University, Maastricht, The Netherlands

5. Department of Methodology & Statistics, Maastricht University, Maastricht, the Netherlands

6. Department of Epidemiology and Biostatistics, Memorial Sloan Kettering Cancer Center, New York, NY, USA

7. Julius Center for Health Sciences and Primary Care, University Medical Center Utrecht, Utrecht University, Utrecht, the Netherlands

8. Respiratory Evaluation Sciences Program, Faculty of Pharmaceutical Sciences, The University of British Columbia, Vancouver, British Columbia, Canada

**Abstract**

External validation studies have finite sample sizes, creating uncertainty about whether a prediction model's Net Benefit (NB) exceeds default strategies' NB. The expected value of perfect information (EVPI) quantifies consequences of uncertainty. Current EVPI methods focus on single studies, ignoring between-center heterogeneity. We extend EVPI and expected value of partial perfect information (EVPPI) to account for between-cluster heterogeneity in multicenter studies and meta-analyses. We distinguish between the global and local optimal strategy and between observed and unobserved clusters. We define $EVPI_{global}$, $EVPI_{cluster\ j}$, $EVPI_{cluster}$, and $EVPPI_{cluster,prevalence}$,, implemented in the MetaNB R package, and illustrate them using a systematic review across 36 centers of the ADNEX model for ovarian cancer diagnosis. Assuming one global decision regarding ADNEX adoption, there is no need for further data to confirm ADNEX is superior overall ($EVPI_{global}$=0). Meta-analysis borrows information across observed clusters, resulting in consistent local superiority of ADNEX and nonzero but typically lower $EVPI_{cluster\ j}$ than when considering local data alone. There is 3% probability default strategies are superior in unobserved centers. Eliminating uncertainty on performance and prevalence in each ($EVPI_{cluster}$) would gain 1134 net avoided false positives (FP) per year, assuming 350000 tumors annually with 20% malignancies. Determining only local prevalence with certainty ($EVPPI_{cluster,\ prevalence}$) would gain net 158 avoided FP per year. EVPI extensions disentangle sources of uncertainty and quantify the need for further validation to determine the global or locally optimal strategy. Considering uncertainty and heterogeneity in clinical utility across clusters is essential to decide whether additional validation studies are warranted.

## Introduction

Clinical prediction models yield a predicted risk or classification into groups corresponding to a diagnosis or prognosis. This statistical or artificial intelligence (AI) prediction can be used by clinicians and patients to support decision-making. For example, a model can predict the risk that a patient's tumor is malignant, and this prediction can be used to select high-risk patients for surgery.

External validation studies estimate model performance in new patients, because the performance of clinical prediction models usually declines when applied in patients that did not contribute data to the model development[1]. These estimates are often uncertain due to sampling variability and typically modest sample sizes. Performance heterogeneity between studies adds a layer of uncertainty. This heterogeneity has many several reasons, including differences in patient spectra[2] and measurement procedures[3]. Ideally, a model should be validated across various settings [4,5] in multicenter studies or through systematic reviews and meta-analyses of validation studies. In this context, the term 'heterogeneity' is used to refer to variance in the underlying true (not observed) performance across clusters (i.e., studies or centers). This variance cannot be reduced by increasing the sample size within clusters, nor the number of clusters we have data on. Between-cluster variance can be estimated, as typically done in random-effects meta-analysis, but estimates are less precise with fewer validations.

If statistical performance or prevalence vary between clusters, we also expect the utility of models to support clinical decision-making to vary.[6,7] A measure of clinical utility that quantifies the benefit of a model for clinical decision-making is Net Benefit (NB).[8] It considers the clinical consequences of using the model to support decisions, not to be confused with Net Benefit in health economics. As a sensitivity analysis, researchers plot the NB of a prediction model and competing strategies across a range of potential predicted risk thresholds to define a high-risk group for which a clinician would deem intervention necessary. This threshold reflects the benefit of a true positive compared to the harm of a false positive. A model with superior NB across the range is considered optimal for supporting decision-making. Decision curve analysis has become an increasingly popular and recommended component of validation studies.[9,10]

Uncertainty quantification for NB has received considerable attention, with some viewing it as a prerequisite for policy change.[11] The focus has been on confidence intervals and hypothesis tests to compare NB between strategies.[12-16] However, requiring statistical significance is inconsistent with the decision-theoretic principle that, when decisions are based on expected outcomes rather than aversion to uncertainty, the strategy with the greatest expected benefit should be adopted (and if necessary, revised later on).[17,18] Nonetheless, uncertainty in NB can result in an incorrect decision to adopt the model, and hence NB loss. The decision-maker might want to evaluate whether further research is required to help increase confidence in their decision. If we had perfect information (infinite validation data), there would not be uncertainty, no risk of an incorrect decision, and no NB loss. The approach of framing the consequences of uncertainty in terms of expected NB loss is referred to as Value of Information (VoI) analysis. We previously defined expected value of perfect information (EVPI) for external validation.[19] However, this EVPI was developed with an external validation in a single setting in mind. Hence, it does not recognize the fact that the utility of a model likely varies between studies or centers.

The aim of the current study is to propose EVPI measures that quantify the expected gain from reducing uncertainty regarding the benefit of a model when the model is validated in multiple centers or studies. In what follows, we first review NB and EVPI for external validation. Next, we define VoI to quantify the expected gain from reducing uncertainty regarding which single strategy would be best globally, and the expected gain from reducing uncertainty regarding which strategy is best in a specific location (e.g., center). For local decisions, we propose VoI measures considering the local performance and disease prevalence are known, and situations where the local prevalence may be known but local performance not. Next, we illustrate the concepts for the decision to recommend use of the diagnostic ADNEX model that estimates the risk of ovarian cancer and guides referral and treatment.

## Methods

### Net Benefit

For a given harm-to-benefit ratio w (equivalently a risk threshold $t$, where $w = t/(1-t)$), Net Benefit (NB) can be expressed in terms of prevalence $\theta^{prev}$, sensitivity $\theta^{sens}$ and specificity $\theta^{spec}$. Net Benefit (NB) for intervening based on a model, intervention in all patients ($NB_{\text{all}}$), and intervention in none ($NB_{\text{none}}$) can be calculated as follows for cluster $j$:[8]

$$\theta_j = \{\theta_j^{prev}, \theta_j^{sens}, \theta_j^{spec}\}$$

$$NB_{model,j}(\theta_j) = \theta_j^{prev} \times \theta_j^{se} - (1 - \theta_j^{prev}) \times (1 - \theta_j^{spec}) \times w$$

$$NB_{all,j}(\theta_j) = \theta_j^{prev} - (1 - \theta_j^{prev}) \times w$$

$$NB_{none,j} = 0$$

For notational convenience, we suppress the dependence on $\theta_j$and write $\text{NB}_{\text{model},j}$, $\text{NB}_{\text{all},j}$. Further details on NB are provided in Appendix I in the Supporting Information.

### Expected Value of Perfect Information in a single validation dataset

The expected value of perfect information corresponds to the difference in NB if the decision-makers had perfect information on the NB of each strategy they consider, and the NB given current uncertain information on each strategy[20]

$$EVPI = \text{NB}_{perfect\ information} - \text{NB}_{current\ information}.$$

The expected gain from eliminating uncertainty is equivalent to the expected loss due to uncertainty surrounding the optimal decision. In the context of a prediction model validated in a single external validation dataset collected in cluster j, the EVPI has previously been introduced as:[20]

$$EVPI_j = \text{E}\{\max(0, NB_{model,j}, NB_{all,j})\} - \max\{0, \text{E}(NB_{model,j}), \text{E}(NB_{all,j})\}$$

We can compute the expectations via bootstrap resampling, although other approaches have been proposed as well.[12,20] The first term is estimated by finding the maximum NB across strategies in each sample and averaging the maxima. The second term is estimated by first averaging NB across samples per strategy, and finding the strategy with the highest expected NB.

## Extending EVPI to multicenter validation and systematic review data

In clustered data, such as multicenter validation data or validation results from multiple populations, prevalence, sensitivity, specificity, $NB_{model}$ and $NB_{all}$ may be different between clusters. Clusters can be physicians, clinics, hospitals or regions, or in the context of systematic review, studies that may or may not coincide with aforementioned examples of clusters. In the context of validation studies of clinical prediction models, we define a *local cluster* as the smallest hierarchically nested data-generating unit within a healthcare system in which the joint distribution of factors relevant to model performance, such as outcome prevalence, patient case-mix, measurement procedures, and clinical workflows, can be considered approximately homogeneous. "Local" is therefore not a fixed geographic or administrative level, but a context-dependent unit determined by the scale at which meaningful heterogeneity affecting prediction performance is expected to arise. When the smallest relevant unit is not observed or not observable, studies or higher-level administrative units may be used as proxies, acknowledging that within-unit heterogeneity may attenuate observed variation in performance. For practical purposes, multiple local units may be aggregated into composite clusters if they are sufficiently similar with respect to key determinants of performance heterogeneity, thereby balancing conceptual fidelity with statistical and logistical feasibility.

Observed performance data from multiple clusters are typically synthesized in a random-effects meta-analysis.[4 21 22] If the meta-analysis is conducted in a Bayesian framework, Markov Chain Monte Carlo (MCMC) sampling methods[23] can naturally be used to sample possible values for uncertain parameters to calculate the posterior distribution of NBs and thus the VoI metrics. We first elaborate on the methods, and then introduce a motivational example.

*Meta-analysis*

In a meta-analytic model for NB proposed previously,[21] a set of parameters $\psi$ describes the distribution of prevalence, sensitivity and specificity across clusters. Assuming the logit sensitivity, logit specificity, and logit prevalence follow a multivariate normal distribution, $\psi$ contains the true mean logit prevalence, true mean logit sensitivity and true mean logit specificity across clusters, as well as their between-cluster variances and covariances [21 24]. Alternatively, a product-normal formulation of the meta-analytic model can be used, in which relations between logit prevalence, logit sensitivity and logit specificity are modelled using a set of linear models [21 24-27]. We elaborate on the meta-analysis model, elements of $\psi$, and their priors in Appendix II.

The result of a Bayesian meta-analysis is a posterior distribution for every estimated quantity, including NB, reflecting our uncertainty after having observed the data. There is uncertainty for each element in $\theta_j$ per cluster due to limited sample sizes per cluster, and for each element in $\psi$ due to the limited number of clusters observed. We can query the posterior distribution to estimate quantities of interest, and generate a predictive distribution to answer questions about new, unobserved clusters.

*NB under current information*

After observing data in cluster j, the decision-maker recommends the local optimal strategy. In the Bayesian meta-analytic setting, we define the expected net benefit with respect to the posterior distribution for cluster $j$, which is influenced by both the data from cluster $j$ and the data from the remaining clusters through the between-cluster model:

$$\mathrm{NB}_{cluster\ j,\ \ current\ information} = \max\left\{\mathrm{E}_{\psi}\mathrm{E}_{\theta_j|\psi}NB_{model,j}, \mathrm{E}_{\psi}\mathrm{E}_{\theta_j|\psi}NB_{all,j},\ \ 0\right\},$$

Which is computed as the maximum NB after averaging the NB of each strategy across posterior draws (Appendix III).

When deciding whether to adopt a model in unobserved clusters j*, the decision-maker may choose one global strategy that has the highest expected NB overall (across randomly selected unobserved clusters). We define the NB under current information, assuming one global strategy, as follows:

$$\mathrm{NB}_{global,\ \ current\ information} = \max\left\{\mathrm{E}_{\psi}\mathrm{E}_{\theta_{j^*}|\psi}NB_{model,j^*},\ \mathrm{E}_{\psi}\mathrm{E}_{\theta_{j^*}|\psi}NB_{all,j^*},\ \ 0\right\}.$$

The computation of this quantity requires sampling from the predictive distribution. This involves two stages (sampling values $\psi$ for parameters of the between-center model, then sampling values for parameters $\theta_{j^*}$ of hypothetical clusters), and taking the maximum across draws, as described in Appendix III.

As $\theta$ varies from one (observed or unobserved) cluster to the next, we remain uncertain due to the limited number of clusters observed, and due to the limited sample size per cluster. The decision-maker can be interested in the value of additional local data collection to determine the strategy that is globally or locally optimal. Hence, the NB under perfect information can be defined in several ways, which we will now explain one by one.

*Perfect information in a specific observed cluster*

The optimal strategy in a specific observed cluster is uncertain due to the limited sample size in that cluster, and the propagated posterior uncertainty in $\psi$ through the meta-analysis model. The NB under perfect information in cluster j is:

$$NB_{cluster\ j,\ \ perfect\ information} = E_{\psi}\, E_{\theta_j|\psi} \max\{NB_{model,j}, NB_{all,j}, 0\},$$

The maximum NB is estimated within each posterior draw (Appendix III). The corresponding EVPI captures the expected NB gain by resolving uncertainties around the NB of the model and competing strategies in cluster j:

$$EVPI_{cluster\ j} = NB_{cluster\ j,\ \ perfect\ information} - NB_{cluster\ j,\ \ current\ information}.$$

Note that by removing uncertainty about the values of $\theta$ in center j, our knowledge about $\psi$ and thus $\theta$ in other centers will also be updated due to the hierarchical nature of the meta-analysis model. However, when the consequences of uncertainty for center j are of concern, it is natural to consider the NB loss due to uncertainty from the perspective of this center only.

*Perfect information at global level*

Some situations may require a global recommendation, such as national guidelines. Decision-makers recommend the global best strategy for unobserved clusters j* (e.g., hospitals) based on samples of finite size of a limited number of observed clusters , and hence we are uncertain whether this strategy is best overall. $\mathrm{EVPI}_{\mathrm{global}}$ captures the expected NB gain if the global strategy is decided with complete information. We define the NB under perfect information, assuming one global strategy, as:

$$NB_{global,\ perfect\ information} = \mathrm{E}_{\psi} \max\left\{E_{\theta_{j^*}|\psi} NB_{model,j^*}, E_{\theta_{j^*}|\psi} NB_{all,j^*}, 0\right\},$$

which is estimated using the two-stage sampling procedure described above, taking the maximum within each posterior draw (Appendix III). The corresponding EVPI is

$$EVPI_{global} = NB_{global,\ perfect\ information} - NB_{global,\ current\ information}.$$

*Perfect information in each unobserved cluster*

Because data are limited to samples of finite size of a limited number of clusters, we are uncertain if the global optimal strategy is optimal in each unobserved cluster. If we know, with certainty, the sensitivity, specificity and prevalence for each cluster, we can identify the best strategy per unobserved cluster. Whereas $NB_{cluster\ j,\ perfect\ information}$ required perfect information only in observed cluster j, we now require perfect information in each unobserved cluster. Whereas $NB_{global,\ perfect\ information}$ assumed one global strategy, this quantity assumes the strategy can vary across currently unobserved clusters. The latter would be more challenging to implement practically as it requires local data collection to inform local choices. The NB under perfect cluster information is:

$$NB_{cluster,\ perfect\ information} = \mathrm{E}_{\psi}\mathrm{E}_{\theta_{j^*}|\psi} \max\left\{NB_{model,j^*}, NB_{all,j^*}, 0\right\},$$

which can be estimated by sampling from the predictive distribution (Appendix III). $\mathrm{EVPI}_{\mathrm{cluster}}$ gives an upper bound for the expected utility gains from local validation to determine (and adopt) the best strategy for each cluster, and is defined as follows:

$$EVPI_{cluster} = NB_{cluster,\ perfect\ information} - NB_{current\ information}.$$

*Partial perfect information in each unobserved cluster*

It may be unfeasible in practice to gather model performance data from each unobserved cluster, and recommending the same strategy everywhere may be suboptimal. Cluster-specific prevalence data may be relatively easy to obtain, and using the optimal strategy for a given prevalence (e.g., treat all in a high-prevalence setting and apply the model elsewhere) may be a good compromise. We define the expected value of patrial perfect information as the value of eliminating uncertainty in local prevalence. The NB under partial perfect[28] cluster information would be:

$$NB_{cluster,\ partial\ perfect\ information} = \mathrm{E}_{\psi}\mathrm{E}_{\theta_{j^*}|\psi} \max\left\{\mathrm{E}_{\theta^c_{j^*}|\theta^i_{j^*},\psi} NB_{model,j^*}, \mathrm{E}_{\theta^c_{j^*}|\theta^i_{j^*},\psi} NB_{all,j^*}, 0\right\},$$

where $\theta^c_{j^*}$ is the prevalence in unobserved clusters and $\theta^c_{j^*}$ is the sensitivity and specificity in unobserved clusters. The NB under partial perfect information is made up of two nested elements. The inner part is the maximum expected NB that would be obtained if uncertainty in prevalence would be resolved. Of course, as the cluster-specific value of prevalence is not known, it is necessary to consider the expectation over the distribution of prevalence, from the meta-analytic model. $NB_{cluster,\ partial\ perfect\ information}$ is estimated using a computationally efficient one-level simulation method explained elsewhere [29], as implemented in the evppi function of the voi R package version 1.0. The values for $NB_{model,j^*}$, $NB_{treat\ all,j^*}$ and $NB_{treat\ none,j^*}$ in sampled clusters, along with their corresponding prevalence are specified as inputs to the function, to calculate the expected value of partial perfect information when eliminating uncertainty in a cluster's prevalence. The corresponding expected value of partial perfect information (EVPPI) is defined as:

$$EVPPI_{cluster,\ prevalence} = NB_{cluster,\ partial\ perfect\ information} - NB_{current\ information}.$$

All EVPI definitions introduced above are available in the MetaNB package (see Appendix IV).

*Motivational example*

We illustrate EVP(P)I for the external validation of ADNEX[30], a clinical risk prediction model that distinguishes between benign and several types of malignant ovarian tumors. Barreñada et al. performed a systematic review of external validations around the world, predominantly Europe and Asia.[31] From these, we use all clusters with available cancer antigen 125; one cluster was excluded as data was confidential. Clusters represent clinics. For studies in which clinic-level data was unavailable, clusters represent study-level data. Some clinics performed multiple validation studies over time, and each instance is included as a separate cluster. The total sample size is 9 889 women, of which 2 984 had a malignant tumor. The observed prevalence varied widely between clusters from 9% to 72%. We investigate the utility of the model to distinguish between benign and malignant tumors. A risk of malignancy $t>10\%$ is recommended to refer women for oncological care.[32] At this threshold, the sensitivity per cluster ranged from 67% to 100%, and specificity from 49% to 94%. We repeat the meta-analysis and EVP(P)I analysis in a subset of 5 clusters to illustrate a scenario where fewer validations have been performed.

We used the JAGS-based MetaNB package in R version 4.5.1 to fit a trivariate meta-analysis model of logit sensitivity, logit specificity and logit prevalence using the product-normal formulation [2]. We used weak realistic priors specified in detail in Appendix V. The model was fitted using the following MCMC parameter settings: two chains, 1000 adaptation iterations, and a burn-in of 3000. Model convergence was assessed using trace plots based on an additional diagnostic sample of 1000 iterations. We drew 2000 triads of logit sensitivity, logit specificity, and logit prevalence from the posterior of $\psi$, representing randomly selected new clusters, to calculate EVP(P)I quantities. The EVPI in a specific observed cluster was calculated in two ways for comparison: using the meta-analytic model, which considers also data from other clusters when estimating unknown parameters for cluster j, and using only data from cluster j.[10]

## Results

The summary estimate for the NB of the ADNEX model (0.27, 95% credible interval (CrI) 0.22 to 0.32, 95% prediction interval (PrI) 0.07 to 0.63) is higher than the NB of treat all (0.23, 95% CrI 0.17 to 0.29, 95% PrI -0.00 to 0.63) and treat none (0). Figure 1 shows that the NB of the performance of the ADNEX model is highly heterogeneous across clusters.

Appendix VI shows standardized NB (NB/prevalence) and the NB difference with the best default strategy per cluster. The 95% prediction interval for the difference includes negative numbers, indicating the model is not the best strategy everywhere (NB difference 0.042, 95% CrI 0.034 to 0.049, 95% PrI -0.002 to 0.077). The posterior probability that ADNEX is superior to treat all and treat none in a randomly selected new cluster (P(useful)) is 0.97.

| Publication | Country | N | Prev | NB | 95% CI |
|---|---|---|---|---|---|
| Van Calster (2020) | UK, Italy | 92 | 10% | 0.05 | [0.00; 0.11] |
| Tug (2020) | Turkey | 285 | 9% | 0.07 | [0.03; 0.10] |
| Joyeux (2016) | France | 284 | 11% | 0.08 | [0.04; 0.11] |
| Van Calster (2020) | Greece | 360 | 18% | 0.12 | [0.08; 0.16] |
| Lam Huong (2022) | Vietnam | 461 | 14% | 0.12 | [0.09; 0.15] |
| Jeong (2020) | South Korea | 54 | 19% | 0.15 | [0.05; 0.25] |
| Poonyakanok (2021) | Thailand | 357 | 17% | 0.16 | [0.12; 0.20] |
| Chen (2022) | Taiwan | 281 | 21% | 0.17 | [0.12; 0.22] |
| Van Calster (2020) | Italy | 121 | 21% | 0.17 | [0.10; 0.24] |
| Van Calster (2020) | Belgium | 211 | 21% | 0.17 | [0.12; 0.23] |
| Viora (2020) | Italy | 577 | 25% | 0.20 | [0.17; 0.24] |
| Van Calster (2020) | Sweden | 278 | 26% | 0.21 | [0.16; 0.26] |
| Lai (2022) | China | 734 | 23% | 0.21 | [0.18; 0.24] |
| Qian (2021) | China | 486 | 25% | 0.21 | [0.17; 0.25] |
| Sandal (2018) | Turkey | 191 | 28% | 0.24 | [0.17; 0.30] |
| Szubert (2016) | Spain | 123 | 28% | 0.25 | [0.17; 0.33] |
| Diaz (2017) | Venezuela | 227 | 30% | 0.26 | [0.21; 0.32] |
| Sayasneh (2016) | UK, Italy | 610 | 30% | 0.27 | [0.23; 0.30] |
| Yang (2022) | China | 376 | 31% | 0.27 | [0.22; 0.32] |
| Van Calster (2020) | Italy | 48 | 31% | 0.28 | [0.15; 0.41] |
| Pelayo (2023) | Spain | 122 | 34% | 0.30 | [0.22; 0.39] |
| Szubert (2016) | Poland | 204 | 34% | 0.30 | [0.24; 0.37] |
| He (2022) | China | 620 | 35% | 0.31 | [0.27; 0.34] |
| Meys (2017) | Netherlands | 326 | 35% | 0.32 | [0.26; 0.37] |
| Zhang (2022) | China | 282 | 37% | 0.32 | [0.26; 0.38] |
| Van Calster (2020) | Poland | 41 | 41% | 0.34 | [0.19; 0.49] |
| Wang And Yang (2023) | China | 445 | 40% | 0.38 | [0.33; 0.42] |
| Van Calster (2020) | Belgium | 202 | 46% | 0.43 | [0.35; 0.50] |
| Peng (2021) | China | 224 | 47% | 0.43 | [0.36; 0.49] |
| Van Calster (2020) | Italy | 354 | 48% | 0.45 | [0.39; 0.50] |
| Van Calster (2020) | Italy | 145 | 50% | 0.46 | [0.37; 0.54] |
| Araujo (2017) | Brazil | 131 | 52% | 0.46 | [0.37; 0.55] |
| Van Calster (2020) | Spain | 54 | 50% | 0.49 | [0.36; 0.63] |
| Van Calster (2020) | Sweden | 239 | 56% | 0.51 | [0.45; 0.58] |
| Van Calster (2020) | Italy | 286 | 56% | 0.54 | [0.48; 0.60] |
| Van Calster (2020) | Italy | 58 | 72% | 0.68 | [0.56; 0.81] |
| Pooled estimate | | | | 0.27 | [0.22; 0.32] |
| Prediction interval | | | | | [0.07; 0.63] |
| P(useful) | | | | 0.97 | |

Figure 1. Meta-analysis results for the ADNEX model

*Note: Squares represent Net Benefit per cluster. Clusters are defined by study, time and location of data collection. The diamond is the 95% Credible Interval around the summary Net Benefit from a random-effects meta-analysis, and the bar represents the 95% prediction interval, which provides a predicted range for the true Net Benefit in a new cluster*

Despite heterogeneity in NB, the global EVPI is 0. There is no loss due to uncertainty regarding the global optimality of ADNEX. If we were to implement one strategy everywhere, there would be no benefit from additional validation studies, and the ADNEX model can safely be recommended.

*Table 1. ADNEX expected value of perfect information*

| | EVPI | Scaled to population<br>Net true positives* | Scaled to population<br>Net avoided false positives* |
|---|---|---|---|
| ***36 external validations*** | | | |
| EVPI $_{global}$ | 0.00000 | 0 | 0 |
| EVPPI $_{cluster, prevalence}$ | 0.00005 | 18 | 158 |
| EVPI $_{cluster}$ | 0.00036 | 126 | 1134 |
| EVPI $_{South-Korea (Jeong 2020)}$ | 0.00001 | <1** | <1** |
| ***5 external validations*** | | | |
| EVPI $_{global}$ | 0.00754 | 2 639 | 2 3751 |
| EVPPI $_{cluster, prevalence}$ | 0.01229 | 4 302 | 3 8714 |
| EVPI $_{cluster}$ | 0.01803 | 6 311 | 5 6795 |

EVPI: Expected value of perfect information; EVPPI: expected value of partial perfect information; *population size 350k assumed; ** population size 1000 assumed.

However, the $EVPPI_{cluster, prevalence}$ and $EVPI_{cluster}$ are >0. This indicates that there are NB gains in eliminating uncertainty regarding the local best strategy, and local studies should be considered. Assuming 350 000 tumors investigated for malignancy in Europe yearly [33-35] the expected gain from eliminating uncertainty regarding the optimal local strategy is a net 126 detected cancers per year, or equivalently a net 1 134 avoided false detections (126 × 0.9/0.1, using the odds of the 10% risk threshold as the exchange rate between true positives and false positives). This would require eliminating uncertainty about sensitivity, specificity and prevalence from each cluster. Only eliminating uncertainty in the local ovarian cancer prevalence and adopting a prevalence-based strategy (treat none for prevalence <0.02, treat all for prevalence >0.65, treat according to model otherwise, Appendix VII) yields an expected gain of a net 18 detected ovarian cancers, or equivalently a net 158 avoided false detections.

Now suppose we only had 5 external validations from Van Calster 2020: Italy (n=48), Poland (n=41), Spain (n=54), Sweden (n=239) and Italy (n=58). The expected gain from reducing uncertainty would be much larger, due to the limited evidence: a net 2 639 detected cancers per year from eliminating uncertainty regarding the best global strategy and a net 6 311 detected cancers per year from eliminating uncertainty regarding the best strategy in each cluster. This indicates a clear need for further validation studies.

Zooming in to a specific cluster, Samsung Medical Center, Seoul, South-Korea (Jeong 2020) the NB of ADNEX and treat all are estimated to be 0.17 and 0.12, respectively, when based on the meta-analysis (Figure 1). Assuming this center performs ultrasound in 1000 patients yearly for suspicion of ovarian cancer, the expected gain from eliminating uncertainty in this center is <1 net true positives and <1 net avoided false positives per year. When using data from this center alone, the estimated NB of ADNEX in this center is 0.14, compared to a NB of treat all of 0.11, with EVPI per 1000 patients translating to 2 net detected ovarian cancers or 18 net avoided false detections per year. Figure 2 illustrates the shrinkage in cluster-specific NB estimates to the overall average when using the meta-analytic model (left panel), and the effect of “borrowing of strength” from data from other clusters on EVPI measures, for all clusters. Based on local data alone, four high-prevalence clusters favor all, but based on meta-analysis, the recommended local strategy is ADNEX for all observed clusters (middle panel). EVPI generally decreases in meta-analysis, compared to using data from only one cluster (right panel). The only exception is the National Cancer Institute in Milan, Italy (Van Calster 2020): the decision to treat all based on 58 local patients with prevalence 0.72 appeared more certain in isolation, than the decision to use ADNEX when local evidence was combined with data from other centers favoring ADNEX.

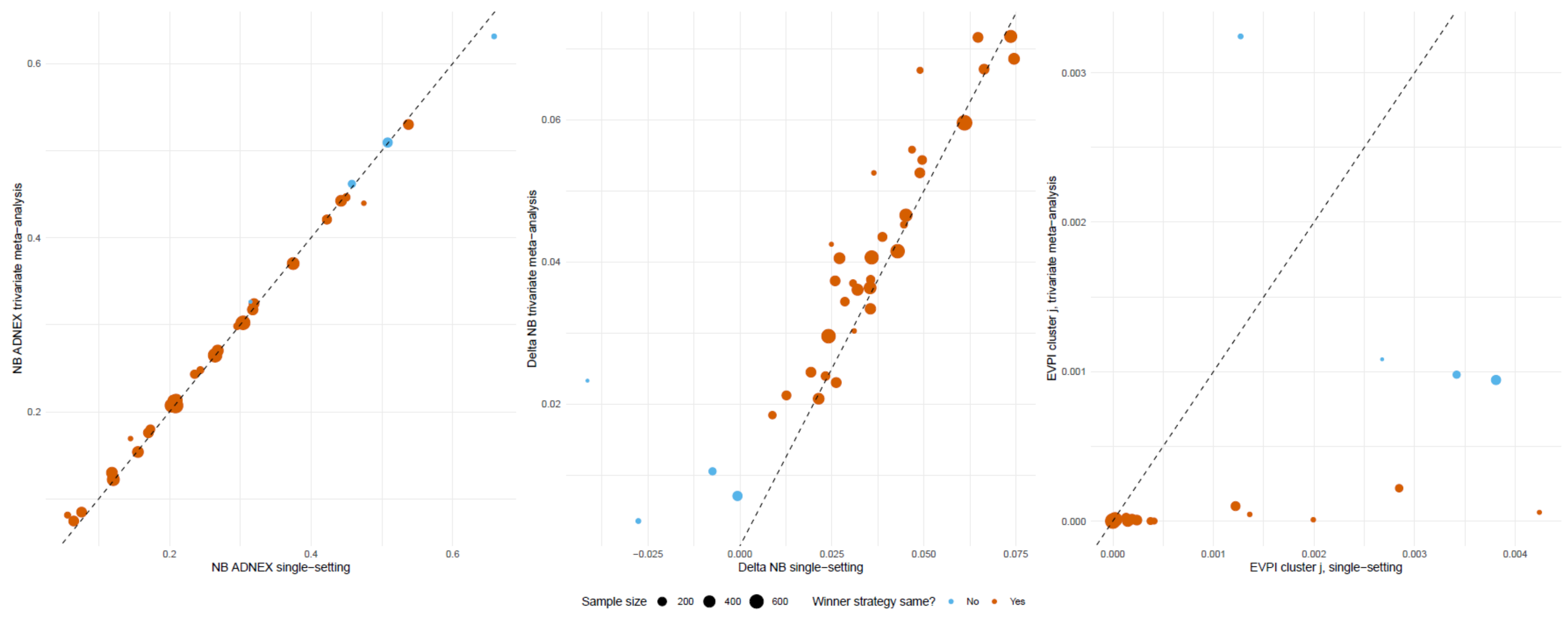

Figure 2: Comparison of results in specific observed clusters when data analysis is performed using only data from that cluster versus when meta-analysis borrows information from other clusters. Left panel: NB of ADNEX, middle panel: NB difference between ADNEX and the best default strategy, right panel: $EVPI_{cluster\ j}$

*Note: The dashed line represents perfect agreement between the two approaches. Point color indicates the winner strategy that gives the highest mean Net Benefit at a threshold of 0.1 in the single cluster analysis, and dot size reflects the sample size in that cluster*

## Discussion

The adoption of clinical prediction models and AI to support medical decision-making should be evidence-based. Ideally, available evidence from validation studies of a particular model is summarized through systematic review and meta-analysis. At the same time, the decision to adopt a model is inevitably taken based on uncertain evidence, since available data originates from a limited number of studied settings (clusters), and each study is limited in size. In the current paper, we proposed a method to estimate the NB and expected value of perfect and partial perfect information in a single analysis. We showed that cluster-specific NB estimates borrow strength and have reduced uncertainty when calculated in a meta-analytic framework. Moreover, we proposed $EVPI_{cluster}$ and $EVPPI_{cluster, prevalence}$ to estimate the NB gain from resolving uncertainty in each unobserved center after eliminating uncertainty on local performance and prevalence, or local prevalence alone, through local data collection in each cluster. Finally, we estimate the gain from further local data collection to eliminate uncertainty about the best global strategy as $EVPI_{global}$.

The uncertainty addressed by our method focuses on epistemic uncertainty: uncertainty that can be addressed with further knowledge. The sources of uncertainty explicitly recognized here are statistical imprecision[36], due to limited sample size per study, and indirectness[36], since data is available on a limited set of clusters, but not all clusters to which the conclusions should be generalizable (e.g., all hospitals in a country). Clusters with direct evidence benefit from indirect evidence from other clusters, instead of only data from that cluster. The reduced uncertainty and local NB estimates shrunken towards overall NB are an example of the classical bias-variance trade-off in Stein's paradox.[37] By introducing a small amount of bias through shrinkage toward the overall mean, the estimator achieves a lower overall mean squared error than estimates based solely on cluster-specific data.[37 38] In other words: to decide on the best strategy on new patients from an already observed cluster j, it is beneficial to borrow information from other clusters, rather than rely completely on the limited observed data from cluster j.

This work aims to contribute to a framework to inform decisions regarding where health care resources should be invested: providing early access to new technologies, versus conducting research to provide additional evidence that can lead to a better understanding of the consequences of using that technology in clinical practice, and the variability in those consequences across settings. To ensure new technologies are implemented and used responsibly, it is important that available evidence and uncertainties are identified and summarized at an early stage. Running many smaller validation studies across the settings of interest may provide more information than a single large "definitive" validation study could ever do. This point has been made over a decade ago for drug trials[39] but bears repeating for new AI technologies in health care, where performance is notoriously context-dependent[40 1 41]

The methods proposed in the current work align with previous recommendations [22 42] to consider predictive distributions (i.e. sampling hypothetical new clusters) and not rely on averages when integrating results from meta-analysis in decision models. It is crucial to consider between-cluster heterogeneity as a source of uncertainty in cost-effectiveness analysis. The $EVPI_{cluster}$ concept introduced in the current paper, which explicitly considers heterogeneity in the optimal decision per cluster, bears resemblance to the concept of value-of-heterogeneity as defined by Espinoza and colleagues for cost-effectiveness analyses,[43] but is not a pure value-of-heterogeneity measure. Espinoza's static value of heterogeneity quantifies the benefit gained from making subgroup-specific decisions, assuming data is observed on all subgroups, while the dynamic value-of-heterogeneity quantifies the expected gain from collecting more data on the observed subgroups to alleviate the uncertainty on subgroup-specific parameters. In contrast, $EVPI_{cluster}$ is a value-of-information measure addressing the fact that some clusters (in practice, the majority), are unobserved, and any inference relies on indirect evidence.

A first limitation of our method is that it cannot reflect uncertainty due to unavailability[36] of information due to a lack of data or insight. For example, clusters included in a meta-analysis may all be from secondary care, but no data on the performance of the model in primary care may be unavailable. In such cases, plausible and extreme scenario analysis may shed some light on the NB in that setting,[36] but additional validation studies will be crucial to inform the decision to implement the model in that setting. Similarly, data on future performance of the model is unavailable by definition. Our method does not replace monitoring of model performance over time.[40 1 41] Monitoring is a regulatory requirement for medical AI devices, although post-deployment performance metrics may not reliably distinguish model deterioration from changes induced by the model itself.[44 45] More work is needed on value of information when monitoring and updating models for changing contexts. Our current approach considers the prediction model as fixed, although models can be updated to improve performance in case of contextual changes over time or when a model is applied in new locations or populations.[46] A second limitation of our work is that it only considers comparisons between the model and default strategies treat all and treat none. While direct comparisons to other strategies in current care or competing AI strategies are straightforward when individual patient data is available, this is not the case when relying on information typically reported in published validation studies. A third limitation is that we have merely quantified the heterogeneity of performance but have not explained it. Future work could investigate an extension to meta-regression, where explanatory factors for differences in utility across clusters can be included. [47] A fourth limitation is it that we have only quantified NB gain under perfect information. To make concrete recommendations on the utility of further validation studies, design and sample size considerations must be considered. Future work will investigate the expected value of sample information and inform decisions to gather more data from many versus a few additional clusters.

In conclusion, our proposed framework quantifies and reduces epistemic uncertainty in clinical prediction model adoption decisions across diverse clinical settings, guiding efficient allocation of research and implementation efforts while highlighting the need for local validation.

**Funding**

This work was supported by the Dutch Research Council (NWO) through the Talent Programme VIDI, project number 09150172310023.

# Supporting Information

## Contents

## Appendix I: Net Benefit

We consider a diagnostic model predicting the risk of a disease being present, although decision curve analysis has been extended for censored prognostic outcomes and counterfactual prediction based on randomized controlled trial data too.[8 48] Patients with a predicted risk above a risk threshold t are classified as positive. We assume that an intervention takes place in patients classified as positive, that is patients with a predicted risk above a given risk threshold t (e.g., invasive diagnostics, treatment). In a validation dataset with N patients, Net Benefit is defined as:

$$NB = \frac{TP - FP \times w}{N},$$

With where TP is the number of true positives and FP the number of false positives. The weight w equals t/(1-t), the odds of the outcome risk t at which the decision-maker has no preference between intervening or not intervening because the expected utility of either option would be equal. Thus, w also corresponds to the harm-to-benefit ratio of a false positive (compared to a true negative) versus a true positive (compared to a false negative), based on an assessment of the intervention given to patients classified as positive. NB has a theoretical range from -∞ to 1 and the highest achievable NB equals the outcome prevalence. The unit of NB is net true positives (i.e. after subtracting a weighted false positive count): NB × 100 gives the net number of true positives per 100 patients. The NB of classifying all patients as negative and not intervening in anyone is 0 by definition. Hence, the NB of a model is to be interpreted as the incremental NB compared to such a "treat none" strategy. Another comparator strategy that can easily be applied in absence of a model is classifying all patients as positive and intervening in everyone ("treat all"). For a model to be useful, it should at least have a higher NB than treat none and treat all. The NB can also be expressed in terms of net FPs avoided by calculating NB / w. Comparison to additional models and strategies is possible too after calculating their NB. Further details of the decision tree underlying the NB formulation, the proof linking threshold and w, and interpretation are given elsewhere.[8] Both poor model discrimination (AUC, ability to discriminate between events and non-events) and poor model calibration (correspondence between predicted risks and observed risks) lower NB. Poorly calibrated models may be clinically harmful, meaning their NB is lower than that of treat all and treat none; the effect of miscalibration depends on the risk threshold t.[7]

## Appendix II: Meta-analysis between-center model formulations and priors

Let $\theta_j^{prev}, \theta_j^{sens}, and\ \theta_j^{spec}$ denote the true prevalence, sensitivity and specificity in cluster j. After logit transformation, these are assumed to follow a multivariate normal distribution:

$$\begin{pmatrix} \text{logit}(\theta_j^{prev}) \\ \text{logit}(\theta_j^{sens}) \\ \text{logit}(\theta_j^{spec}) \end{pmatrix} \sim \text{N}\left[\begin{pmatrix} \gamma_1 \\ \gamma_2 \\ \gamma_3 \end{pmatrix}, \mathbf{\Omega}\right], \mathbf{\Omega} = \begin{pmatrix} \tau_1^2 & \rho_{12}\tau_1\tau_2 & \rho_{13}\tau_1\tau_3 \\ \rho_{12}\tau_1\tau_2 & \tau_2^2 & \rho_{23}\tau_2\tau_3 \\ \rho_{13}\tau_1\tau_3 & \rho_{23}\tau_2\tau_3 & \tau_3^2 \end{pmatrix}$$

where $\gamma_1, \gamma_2, and\ \gamma_3$ are the summary logit prevalence, logit sensitivity, and logit specificity, and $\mathbf{\Omega}$ is the between-cluster variance-covariance matrix with variances $\tau_1^2, \tau_2^2, and\ \tau_3^2$ and pairwise correlations $\rho_{12}, \rho_{13}, and\ \rho_{23}$.

Often, normal priors for the means and Wishart prior for the covariance matrix are used when performing a Bayesian multivariate meta-analysis. The Wishart prior places a joint distribution on the entire covariance matrix, where a scale matrix and degrees of freedom simultaneously determine priors for all parameters of substantive interest (variances and pairwise correlations). Hence, this prior does not allow you to specify prior beliefs separately for variance and correlation parameters. Many meta-analyses include relatively few studies. In such situations, posterior estimates may be sensitive to the prior specification, and the Wishart prior may be unexpectedly and unintendingly informative. This motivated different between-center model formulations.[25-27]

In the product-normal formulation, the multivariate normal between-cluster model is reparametrized as a sequence of linear models: [21 24-27]

$$\begin{cases} \text{logit}(\theta_j^{prev}) \sim \text{N}(\eta_1, \sigma_1^2) \\ \text{logit}(\theta_j^{sens})|\text{logit}(\theta_j^{prev}) \sim \text{N}(\eta_{2\text{j}}, \sigma_2^2) \\ \eta_{2\text{j}} = \lambda_{20} + \lambda_{21}\text{logit}(\theta_j^{prev}) \\ \text{logit}(\theta_j^{spec})|\text{logit}(\theta_j^{prev}), \text{logit}(\theta_j^{sens}) \sim \text{N}(\eta_{3\text{j}}, \sigma_3^2) \\ \eta_{3\text{j}} = \lambda_{30} + \lambda_{31}\text{logit}(\theta_j^{prev}) + \lambda_{32}\text{logit}(\theta_j^{sens}) \end{cases}$$

The conditional variances $\sigma_1^2, \sigma_1^2, and\ \sigma_1^2$ and coefficients $\lambda_{21}, \lambda_{31}, and\ \lambda_{32}$ are related to the elements of $\mathbf{\Omega}$ by:

$$\sigma_1^2 = \tau_1^2, \qquad \sigma_2^2 = \tau_2^2 - \lambda_{21}^2\tau_1^2, \qquad \sigma_3^2 = \tau_3^2 - \lambda_{31}^2\tau_1^2 - \lambda_{32}^2\tau_2^2 - 2\lambda_{31}\lambda_{32}\lambda_{21}\tau_1^2$$

$$\lambda_{21} = \frac{\tau_2}{\tau_1}\rho_{12}, \qquad \lambda_{31} = \frac{\tau_3(\rho_{13} - \rho_{12}\rho_{23})}{\tau_1(1 - \rho_{12}^2)}, \qquad \lambda_{32} = \frac{\tau_3(\rho_{23} - \rho_{12}\rho_{13})}{\tau_2(1 - \rho_{12}^2)}$$

This allows separate priors to be placed on $\tau_1^2, \tau_2^2, and\ \tau_3^2$ and $\rho_{12}, \rho_{13}, and\ \rho_{23}$ directly, rather than on $\mathbf{\Omega}$ as a whole. The priors used in the case study are described in Appendix II. The summary estimates are recovered as:

$$\gamma_1 = \eta_1, \qquad \gamma_2 = \lambda_{20} + \lambda_{21}\gamma_1, \qquad \gamma_3 = \lambda_{30} + \lambda_{31}\gamma_1 + \lambda_{32}\gamma_2$$

In the main paper, we group the parameters $\Upsilon$, $\rho$, and $\tau$ into a single vector $\psi$ for ease of exposition.

## Appendix III: MCMC algorithms

**Algorithm for estimating $\text{NB}_{cluster\ j,\ \ current\ information}$ (i.e., for observed cluster j)**

For $m = 1, \dots, M$ posterior draws:

1. Obtain a posterior draw of the cluster-specific parameters $\theta_j^{(m)} \sim p(\theta_j | D)$ where $p(\theta_j \mid D)$ denotes the marginal posterior distribution of $\theta_j$ under the meta-analysis model.
2. Calculate $\text{NB}_{\text{model},j}^{(m)} = \text{NB}_{\text{model}}(\theta_j^{(m)})$ and $\text{NB}_{\text{all},j}^{(m)} = \text{NB}_{\text{all}}(\theta_j^{(m)})$.
3. Store $\text{NB}_{\text{model},j}^{(m)}$ and $\text{NB}_{\text{all},j}^{(m)}$.
4. After all posterior draws have been processed, calculate $\widehat{\text{ENB}}_{\text{model, j}} = \frac{1}{M}\sum_{m=1}^{M} \text{NB}_{\text{model},j}^{(m)}$ and $\widehat{\text{ENB}}_{\text{all, j}} = \frac{1}{M}\sum_{m=1}^{M} \text{NB}_{\text{all},j}^{(m)}$. *
5. Finally, estimate $\widehat{\text{NB}}_{cluster\ j,\ \ current\ information} = \max\{\widehat{\text{ENB}}_{\text{model, j}}, \widehat{\text{ENB}}_{\text{all, j}}, 0\}$.

* These quantities estimate $\text{E}_{\psi}\text{E}_{\theta_j|\psi}NB_{model,j}$ and $\text{E}_{\psi}\text{E}_{\theta_j|\psi}NB_{all,j}$, respectively, because the marginal posterior $p(\theta_j \mid D)$integrates over uncertainty in $\psi$.

**Algorithm for estimating $\text{NB}_{global,\ \ current\ information}$**

For $m = 1, \dots, M$ posterior draws:

1. Sample the parameters of the between-cluster model $\psi^{(m)} \sim p(\psi|D)$
2. Given the between-cluster distribution specified by $\psi^{(m)}$, sample cluster-specific parameters for a hypothetical new cluster $\theta_{j^*}^{(m)} \sim p(\theta \mid \psi^{(m)})$.
3. Calculate and store $\text{NB}_{\text{model},j^*}^{(m)} = \text{NB}_{\text{model}}(\theta_{j^*}^{(m)})$ and $\text{NB}_{\text{all},j^*}^{(m)} = \text{NB}_{\text{all}}(\theta_{j^*}^{(m)})$.
4. After all posterior draws have been processed, calculate $\widehat{\text{ENB}}_{\text{model}} = \frac{1}{M}\sum_{m=1}^{M} \text{NB}_{\text{model},j^*}^{(m)}$ and $\widehat{\text{ENB}}_{\text{all}} = \frac{1}{M}\sum_{m=1}^{M} \text{NB}_{\text{all},j^*}^{(m)}$.
5. Finally, estimate $\widehat{\text{NB}}_{global,\ \ current\ information} = \max\{\widehat{\text{ENB}}_{\text{model}}, \widehat{\text{ENB}}_{\text{all}}, 0\}$.

**Algorithm for estimating $\text{NB}_{cluster\ j,\ \ perfect\ information}$ (i.e., for a specific observed cluster j)**

For $m = 1, \dots, M$ posterior draws:

1. Obtain a posterior draw of the cluster-specific parameters $\theta_j^{(m)} \sim p(\theta_j | D)$ where $p(\theta_j \mid D)$ denotes the marginal posterior distribution of $\theta_j$ under the meta-analysis model.
2. Calculate $\text{NB}_{\text{model},j}^{(m)} = \text{NB}_{\text{model}}(\theta_j^{(m)})$, $\text{NB}_{\text{all},j}^{(m)} = \text{NB}_{\text{all}}(\theta_j^{(m)})$, and $\text{NB}_{\text{cluster j, perfect information}}^{(m)} = \max\{\text{NB}_{\text{model}}^{(m)}, \text{NB}_{\text{all}}^{(m)}, 0\}$.
3. Store $\text{NB}_{\text{cluster j, perfect information}}^{(m)}$
4. After all posterior draws have been processed, calculate $\widehat{\text{NB}}_{\text{cluster j, perfect information}} = \frac{1}{M}\sum_{m=1}^{M} \text{NB}_{\text{cluster j, perfect information}}^{(m)}$ *

* This quantity estimates $E_{\psi}\, E_{\theta_j|\psi} \max\{NB_{model,j}, NB_{all,j}, 0\}$, because the marginal posterior $p(\theta_j \mid D)$integrates over uncertainty in $\psi$.

**Algorithm for estimating NB$_{global,\ perfect\ information}$**

For $m = 1, \dots, M$ posterior draws:

1. Sample the parameters of the between-cluster model $\psi^{(m)} \sim p(\psi|D)$
2. For $j^* = 1, \dots, J^*$ hypothetical clusters:
   a. Given the between-cluster distribution specified by $\psi^{(m)}$, sample cluster-specific parameters $\theta_{j^*}^{(m)} \sim p(\theta \mid \psi^{(m)})$.
   b. Calculate $\mathrm{NB}_{\mathrm{model},j^*}^{(m)} = \mathrm{NB}_{\mathrm{model}}(\theta_{j^*}^{(m)})$ and $\mathrm{NB}_{\mathrm{all},j^*}^{(m)} = \mathrm{NB}_{\mathrm{all}}(\theta_{j^*}^{(m)})$.
3. Calculate the average net benefit across hypothetical clusters:
$$\bar{\mathrm{NB}}_{\mathrm{model}}^{(m)} = \frac{1}{J^*}\sum_{j^*=1}^{J^*} \mathrm{NB}_{\mathrm{model},j^*}^{(m)},$$
$$\bar{\mathrm{NB}}_{\mathrm{all}}^{(m)} = \frac{1}{J^*}\sum_{j^*=1}^{J^*} \mathrm{NB}_{\mathrm{all},j^*}^{(m)}.$$
4. Under perfect information about $\theta$ given $\psi^{(m)}$, determine the optimal global strategy:
$$\mathrm{NB}_{\text{global, perfect information}}^{(m)} = \max\left\{\bar{\mathrm{NB}}_{\mathrm{model}}^{(m)}, \bar{\mathrm{NB}}_{\mathrm{all}}^{(m)}, 0\right\}.$$
5. Store $\mathrm{NB}_{\text{global, perfect information}}^{(m)}$.
6. After all posterior draws, estimate expected net benefit under perfect information:
$$\widehat{\mathrm{NB}}_{\text{global, perfect information}} = \frac{1}{M}\sum_{m=1}^{M} \mathrm{NB}_{\text{global, perfect}}^{(m)}.$$

**Algorithm for estimating NB$_{cluster,\ perfect\ information}$ (i.e., for each unobserved cluster j*)**

For $m = 1, \dots, M$ posterior draws:

1. Sample the parameters of the between-cluster model $\psi^{(m)} \sim p(\psi|D)$
2. Given the between-cluster distribution specified by $\psi^{(m)}$, sample cluster-specific parameters $\theta_{j^*}^{(m)} \sim p\big(\theta \mid \psi^{(m)}\big)$.
3. Calculate $\mathrm{NB}_{\mathrm{model},j^*}^{(m)} = \mathrm{NB}_{\mathrm{model}}(\theta_{j^*}^{(m)})$ and $\mathrm{NB}_{\mathrm{all},j^*}^{(m)} = \mathrm{NB}_{\mathrm{all}}(\theta_{j^*}^{(m)})$.
4. Determine and store $\mathrm{NB}_{\text{cluster } j^*\text{, perfect information}}^{(m)} = \max\left\{\mathrm{NB}_{\mathrm{model}}^{(m)}, \mathrm{NB}_{\mathrm{all}}^{(m)}, 0\right\}$.
5. After all posterior draws, estimate expected net benefit under perfect information:

$$\widehat{\text{NB}}_{\text{cluster, perfect information}} = \frac{1}{M}\sum_{m=1}^{M} \text{NB}^{(m)}_{\text{cluster } j^*\text{, perfect information}}.$$

## Appendix IV: R code

The package MetaNB to conduct meta-analysis and value of information analysis is available from https://github.com/zhipeiwang/MetaNB. In R, the package can be installed through the following commands :

```
remotes::install_github("zhipeiwang/MetaNB")

library(MetaNB)
```

The vignette available on GitHub provides a tutorial to conduct meta-analysis and calculate EVP(P)I measures.

## Appendix V: Case study priors

We used the vague realistic priors recommended in Wynants et al. [18] Vague normal priors were placed on $\eta_1, \lambda_{20}\ and\ \lambda_{30} \sim\ \mathrm{N}(0, 1000)$. Half-normal priors were used for the between-cluster standard variances, $\tau_1^2, \tau_1^2, and\ \tau_1^2 \sim\ \mathrm{N}(0, 4)$ I(0). For the between-cluster correlations, a uniform prior $\mathrm{U}[-0.99, 0.99]$ was used for $\rho_{12}$, the correlation between logit sensitivity and logit prevalence. Fisher-transformed priors $z \sim\ \mathrm{N}(-0.20, 0.25)$, where $z\ = log[(1 + \rho)/(1 - \rho)]$, were used for the correlations between logit sensitivity and logit specificity $\rho_{23}$ and logit specificity and logit prevalence $\rho_{13}$.

## Appendix VI: Standardized NB and $NB_{model}$-max($NB_{all}$, 0)

| Publication | Country | N | Prev | sNB | 95% CI |
|---|---|---|---|---|---|
| Van Calster (2020) | UK, Italy | 92 | 10% | 0.56 | [0.03; 1.09] |
| Van Calster (2020) | Greece | 360 | 18% | 0.66 | [0.46; 0.86] |
| Tug (2020) | Turkey | 285 | 9% | 0.71 | [0.36; 1.07] |
| Joyeux (2016) | France | 284 | 11% | 0.72 | [0.39; 1.05] |
| Van Calster (2020) | Sweden | 278 | 26% | 0.80 | [0.60; 0.99] |
| Jeong (2020) | South Korea | 54 | 19% | 0.80 | [0.25; 1.35] |
| Van Calster (2020) | Poland | 41 | 41% | 0.81 | [0.45; 1.17] |
| Viora (2020) | Italy | 577 | 25% | 0.82 | [0.68; 0.96] |
| Chen (2022) | Taiwan | 281 | 21% | 0.83 | [0.60; 1.06] |
| Van Calster (2020) | Italy | 121 | 21% | 0.84 | [0.50; 1.17] |
| Van Calster (2020) | Belgium | 211 | 21% | 0.84 | [0.58; 1.10] |
| Qian (2021) | China | 486 | 25% | 0.85 | [0.69; 1.00] |
| Sandal (2018) | Turkey | 191 | 28% | 0.86 | [0.62; 1.09] |
| Lam Huong (2022) | Vietnam | 461 | 14% | 0.86 | [0.64; 1.08] |
| Yang (2022) | China | 376 | 31% | 0.86 | [0.71; 1.02] |
| He (2022) | China | 620 | 35% | 0.87 | [0.76; 0.97] |
| Zhang (2022) | China | 282 | 37% | 0.87 | [0.71; 1.03] |
| Diaz (2017) | Venezuela | 227 | 30% | 0.88 | [0.69; 1.08] |
| Szubert (2016) | Poland | 204 | 34% | 0.88 | [0.69; 1.08] |
| Sayasneh (2016) | UK, Italy | 610 | 30% | 0.89 | [0.76; 1.01] |
| Pelayo (2023) | Spain | 122 | 34% | 0.89 | [0.64; 1.15] |
| Araujo (2017) | Brazil | 131 | 52% | 0.90 | [0.72; 1.07] |
| Van Calster (2020) | Italy | 48 | 31% | 0.90 | [0.48; 1.32] |
| Lai (2022) | China | 734 | 23% | 0.90 | [0.77; 1.03] |
| Szubert (2016) | Spain | 123 | 28% | 0.90 | [0.61; 1.19] |
| Meys (2017) | Netherlands | 326 | 35% | 0.90 | [0.75; 1.06] |
| Peng (2021) | China | 224 | 47% | 0.91 | [0.77; 1.05] |
| Van Calster (2020) | Sweden | 239 | 56% | 0.91 | [0.79; 1.03] |
| Poonyakanok (2021) | Thailand | 357 | 17% | 0.91 | [0.68; 1.15] |
| Van Calster (2020) | Italy | 145 | 50% | 0.92 | [0.75; 1.09] |
| Van Calster (2020) | Belgium | 202 | 46% | 0.92 | [0.77; 1.08] |
| Van Calster (2020) | Italy | 354 | 48% | 0.93 | [0.81; 1.04] |
| Wang And Yang (2023) | China | 445 | 40% | 0.93 | [0.82; 1.04] |
| Van Calster (2020) | Italy | 58 | 72% | 0.94 | [0.78; 1.11] |
| Van Calster (2020) | Italy | 286 | 56% | 0.97 | [0.86; 1.08] |
| Van Calster (2020) | Spain | 54 | 50% | 0.98 | [0.71; 1.25] |
| Pooled estimate | | | | 0.88 | [0.86; 0.90] |
| Prediction interval | | | | | [0.66; 0.96] |

| Publication | Country | N | Prev | dNB | 95% CI |
|---|---|---|---|---|---|
| Van Calster (2020) | Poland | 41 | 41% | -0.01 | [-0.16; 0.14] |
| Van Calster (2020) | Italy | 58 | 72% | -0.01 | [-0.13; 0.11] |
| Araujo (2017) | Brazil | 131 | 52% | -0.00 | [-0.09; 0.09] |
| Van Calster (2020) | Sweden | 239 | 56% | 0.00 | [-0.07; 0.07] |
| Van Calster (2020) | Italy | 145 | 50% | 0.02 | [-0.07; 0.10] |
| Peng (2021) | China | 224 | 47% | 0.02 | [-0.05; 0.08] |
| Zhang (2022) | China | 282 | 37% | 0.02 | [-0.04; 0.08] |
| Van Calster (2020) | Italy | 354 | 48% | 0.02 | [-0.03; 0.08] |
| Van Calster (2020) | Belgium | 202 | 46% | 0.03 | [-0.05; 0.10] |
| He (2022) | China | 620 | 35% | 0.03 | [-0.01; 0.06] |
| Van Calster (2020) | Sweden | 278 | 26% | 0.03 | [-0.02; 0.08] |
| Van Calster (2020) | Greece | 360 | 18% | 0.03 | [-0.01; 0.07] |
| Van Calster (2020) | Italy | 286 | 56% | 0.03 | [-0.03; 0.09] |
| Szubert (2016) | Poland | 204 | 34% | 0.03 | [-0.03; 0.10] |
| Yang (2022) | China | 376 | 31% | 0.03 | [-0.01; 0.08] |
| Wang And Yang (2023) | China | 445 | 40% | 0.04 | [-0.01; 0.08] |
| Meys (2017) | Netherlands | 326 | 35% | 0.04 | [-0.02; 0.09] |
| Viora (2020) | Italy | 577 | 25% | 0.04 | [0.00; 0.07] |
| Pelayo (2023) | Spain | 122 | 34% | 0.04 | [-0.05; 0.12] |
| Sandal (2018) | Turkey | 191 | 28% | 0.04 | [-0.02; 0.11] |
| Diaz (2017) | Venezuela | 227 | 30% | 0.04 | [-0.02; 0.10] |
| Van Calster (2020) | Italy | 48 | 31% | 0.04 | [-0.09; 0.18] |
| Sayasneh (2016) | UK, Italy | 610 | 30% | 0.04 | [0.01; 0.08] |
| Qian (2021) | China | 486 | 25% | 0.05 | [0.01; 0.08] |
| Van Calster (2020) | Spain | 54 | 50% | 0.05 | [-0.09; 0.18] |
| Szubert (2016) | Spain | 123 | 28% | 0.05 | [-0.03; 0.13] |
| Jeong (2020) | South Korea | 54 | 19% | 0.05 | [-0.05; 0.16] |
| Chen (2022) | Taiwan | 281 | 21% | 0.05 | [0.01; 0.10] |
| Van Calster (2020) | Italy | 121 | 21% | 0.05 | [-0.02; 0.12] |
| Van Calster (2020) | Belgium | 211 | 21% | 0.05 | [0.00; 0.11] |
| Van Calster (2020) | UK, Italy | 92 | 10% | 0.05 | [0.00; 0.11] |
| Lai (2022) | China | 734 | 23% | 0.06 | [0.03; 0.09] |
| Tug (2020) | Turkey | 285 | 9% | 0.07 | [0.03; 0.10] |
| Joyeux (2016) | France | 284 | 11% | 0.07 | [0.03; 0.11] |
| Lam Huong (2022) | Vietnam | 461 | 14% | 0.08 | [0.04; 0.11] |
| Poonyakanok (2021) | Thailand | 357 | 17% | 0.08 | [0.04; 0.12] |
| Pooled estimate | | | | 0.04 | [0.03; 0.05] |
| Prediction interval | | | | | [-0.00; 0.07] |

-0.1 0 0.1
Delta Net Benefit

"Delta Net Benefit" or dNB=$NB_{model}$-max($NB_{all}$, 0). Center-specific estimates are obtained from reported sensitivity, specificity and prevalence in primary studies using the formula (TP – FP * w)/N – max(0, (prev*N – (1-prev)*N*w)/N), with TP the number of true positives, FP the number of false positives, prev the prevalence and w=0.1/0.9. Note that Van Calster (2020) Sweden has dNB of 0.0002 here while its corresponding dNB in Figure 2 is -0.0007.

dNB in Figure 2 is the result of a fully Bayesian analysis[12] with uniform priors, which explains differences in point estimates.

## Appendix VII: Optimal strategy conditional on prevalence

| prev_known | NB_TN_mean | NB_model_mean | NB_TA mean | probuseful[a] | winner_strategy[b] |
|---|---|---|---|---|---|
| 0.01 | 0 | -0.0006 | -0.1000 | 0.5490 | treat_none |
| 0.02 | 0 | 0.0052 | -0.0889 | 0.8325 | model |
| 0.03 | 0 | 0.0123 | -0.0778 | 0.9485 | model |
| 0.04 | 0 | 0.0199 | -0.0667 | 0.9800 | model |
| 0.05 | 0 | 0.0279 | -0.0556 | 0.9935 | model |
| 0.15 | 0 | 0.1170 | 0.0556 | 1.0000 | model |
| 0.25 | 0 | 0.2125 | 0.1667 | 0.9960 | model |
| 0.35 | 0 | 0.3107 | 0.2778 | 0.9780 | model |
| 0.45 | 0 | 0.4110 | 0.3889 | 0.9480 | model |
| 0.55 | 0 | 0.5128 | 0.5000 | 0.8710 | model |
| 0.56 | 0 | 0.5230 | 0.5111 | 0.8635 | model |
| 0.57 | 0 | 0.5333 | 0.5222 | 0.8515 | model |
| 0.58 | 0 | 0.5436 | 0.5333 | 0.8395 | model |
| 0.59 | 0 | 0.5539 | 0.5444 | 0.8280 | model |
| 0.60 | 0 | 0.5642 | 0.5556 | 0.8165 | model |
| 0.65 | 0 | 0.6160 | 0.6111 | 0.7410 | model |
| 0.75 | 0 | 0.7209 | 0.7222 | 0.5490 | treat_all |
| 0.85 | 0 | 0.8276 | 0.8333 | 0.3390 | treat_all |
| 0.90 | 0 | 0.8821 | 0.8889 | 0.2310 | treat_all |

Abbreviations: prev_known, known prevalence; NB, net benefit; TN, treat none; TA, treat all; [a]probuseful, probability of usefullness in a new cluster, determined as the probability that ADNEX is superior to treat all and treat none, based on the (wide and skewed) posterior predictive distribution of NB for ADNEX, treat all, and 0, representing NB values in new clusters. Due to skewness, the strategy with the highest probuseful does not necessarily have the highest expected value. [b]The winning strategy is based on the mean NB across new clusters, calculated from the predictive distribution.